\documentclass[letterpaper, 10 pt, conference]{ieeeconf}  

\IEEEoverridecommandlockouts                              

\usepackage{graphics} 
\usepackage{epsfig} 
\usepackage{mathptmx} 
\usepackage{times} 
\usepackage{amsmath} 
\usepackage{amssymb}  
\usepackage{color}

\usepackage{hyperref}

\newtheorem{theorem}{Theorem}[section]

\newtheorem{corollary}{Corollary}[section]

\newtheorem{definition}{Definition}[section]
\newtheorem{example}{Example}[section]
\newtheorem{exercise}{Exercise}[section]
\newtheorem{lemma}{Lemma}[section]

\newtheorem{problem}{Problem}[section]
\newtheorem{proposition}{Proposition}[section]
\newtheorem{remark}{Remark}[section]
\newtheorem{assumption}{Assumption}[section]
\newcommand{\bthm}{\begin{theorem}}
\newcommand{\ethm}{\end{theorem}}
\newcommand{\blem}{\begin{lemma}}
\newcommand{\elem}{\end{lemma}}
\newcommand{\bex}{\begin{example}}
\newcommand{\eex}{\end{example}}
\newcommand{\beg}{\begin{exercise}}
\newcommand{\eeg}{\end{exercise}}
\newcommand{\bprop}{\begin{proposition}}
\newcommand{\eprop}{\end{proposition}}
\newcommand{\bplm}{\begin{problem}}
\newcommand{\eplm}{\end{problem}}
\newcommand{\bmrk}{\begin{remark}}
\newcommand{\emrk}{\end{remark}}
\newcommand{\bdfn}{\begin{definition}}
\newcommand{\edfn}{\end{definition}}
\newcommand{\bcor}{\begin{corollary}}
\newcommand{\ecor}{\end{corollary}}

\newcommand{\beq}{\begin{equation}}
\newcommand{\eeq}{\end{equation}}
\newcommand{\beqm}{\begin{equation*}}
\newcommand{\eeqm}{\end{equation*}}
\newcommand{\beqn}{\begin{eqnarray}}
\newcommand{\eeqn}{\end{eqnarray}}
\newcommand{\beqnm}{\begin{eqnarray*}}
\newcommand{\eeqnm}{\end{eqnarray*}}
\newcommand{\bea}{\begin{aligned}}
\newcommand{\eea}{\end{aligned}}
\newcommand{\beam}{\begin{aligned*}}
\newcommand{\eeam}{\end{aligned*}}
\newcommand{\bs}{\begin{subequations}}
\newcommand{\es}{\end{subequations}}

\newcommand{\bei}{\begin{itemize}}
\newcommand{\eei}{\end{itemize}}
\newcommand{\bed}{\begin{description}}
\newcommand{\eed}{\end{description}}
\newcommand{\bee}{\begin{enumerate}}
\newcommand{\eee}{\end{enumerate}}
\newcommand{\bey}{\begin{array}}
\newcommand{\eey}{\end{array}}
\newcommand{\bec}{\begin{center}}
\newcommand{\eec}{\end{center}}
\newcommand{\la}{\label}

\newcommand{\mbf}{\mathbf}

\title{\LARGE \bf
On Poles and Zeros of Linear Quantum Systems
\thanks{This work is partially financially supported by Innovation Program for Quantum Science and Technology 2023ZD0300600, Guangdong Provincial Quantum Science Strategic Initiative (No. GDZX2200001), Hong Kong Research Grant Council (RGC) under Grant No. 15213924, National Natural Science Foundation of China under Grants Nos. 62173288, 62003111, Natural Science Foundation of Guangdong Province under Grant No. 2022A1515010390, and The Science Center Program of National Natural Science Foundation of China under Grant No. 62188101. (Corresponding author: Guofeng Zhang.)}
}

\author{\parbox{3 in}{\centering Zhiyuan Dong$^\dagger$
        \thanks{$^\dagger$Zhiyuan Dong is with School of Science, Harbin Institute of Technology, Shenzhen, China 
        {\tt\small dongzhiyuan@hit.edu.cn}}}
        \hspace*{- 1.5 in}
        \parbox{3 in}{ \centering Guofeng Zhang$^\sharp$
        \thanks{$^\sharp$Guofeng Zhang is with Department of Applied Mathematics, The Hong Kong Polytechnic University, Hong Kong and also with Shenzhen Research Institute, The Hong Kong Polytechnic University, Shenzhen, China 
        {\tt\small guofeng.zhang@polyu.edu.hk}}}
        \hspace*{- 1.5 in}
        \parbox{3 in}{ \centering Heung-wing Joseph Lee$^\S$
        \thanks{$^\S$Heung-wing Joseph Lee is with Department of Applied Mathematics, The Hong Kong Polytechnic University, Hong Kong, China
        {\tt\small joseph.lee@polyu.edu.hk}}}
}

\begin{document}

\maketitle
\thispagestyle{empty}
\pagestyle{empty}

\begin{abstract}

The non-commutative nature of quantum mechanics imposes fundamental constraints on system dynamics, which in the linear realm are manifested by the physical realizability conditions on system matrices. These restrictions endow system matrices with special structure. The purpose of this paper is to study such structure by investigating  zeros and poses of linear quantum systems. In particular, we show that $-s_0^\ast$ is a transmission zero if and only if $s_0$ is a pole, and which is further generalized to the relationship between system eigenvalues and invariant zeros. Additionally, we study left-invertibility and fundamental tradeoff for linear quantum systems in terms of their zeros and poles.

\end{abstract}

\section{Introduction}

In systems and control theory, zeros and poles are important concepts  which play a significant role in the dynamics and control design of linear dynamical systems. There is a wide range of definitions of system zeros, including decoupling zero \cite{HJ84}, block zero \cite{FB77,Patel86,CHEN1992}, transmission zero and invariant zero \cite{kailath1980,DCC1982,CD1991,ZDG96}.   Simply speaking, transmission zeros represent the frequencies at which the system's output is zero, regardless of the input. In other words, transmission zeros are the frequencies at which the system's transmission function becomes zero. Non-minimum phase (namely zeros on the  right half of the complex plane) and unstable poles often pose fundamental performance tradeoffs in control system design  \cite{bode1945network,SP05,SBG12,CI19}.   Recently, the left-invertibility of linear systems is investigated in \cite{DD2023} by means of invariant zeros, where a constructive procedure for the input reconstruction is proposed.  

In recent decades, significant advancements have been made in both theoretical understanding and experimental applications of quantum control. Quantum control plays a pivotal role in various quantum information technologies, such as quantum communication, quantum computation, quantum cryptography, quantum ultra-precision metrology, and nano-electronics. Similar to classical control systems theory, linear quantum systems hold great importance in the field of quantum control. Linear quantum systems are mathematical models that describe the behavior of quantum harmonic oscillators. In this context, ``linear'' refers to the linearity of the Heisenberg equations of motion for quadrature operators in the quantum system. This linearity often leads to simplifications that facilitate analysis and control of these systems. Consequently, linear quantum systems can be effectively studied using powerful mathematical techniques derived from linear systems theory. A wide range of quantum-mechanical systems can be suitably modeled as linear quantum systems. For instance, quantum optical
systems, 
circuit quantum electro-dynamical (circuit QED) systems, 
cavity QED systems,
 quantum opto-mechanical
systems, 
atomic ensembles,
and
quantum memories.


It is very plausible to speculate that poles and  zeros are important in  linear quantum systems theory \cite{YJ14,ZD22}. For example, as shown in  \cite{Yan07},  these concepts are fundamental to understanding the behavior and characteristics of linear quantum  systems. This paper aims to delve into the intricacies of invariant zeros, transmission zeros, invertibility, and sensitivity of linear quantum systems, elucidating their roles, relationships, and implications in quantum control theory. We show that the concept of invariant zeros in linear quantum systems is closely related to the eigenvalues of system matrix. Meanwhile, there exists a one-to-one correspondence between system poles and transmission zeros. Understanding the locations and characteristics of zeros and poles is crucial for analyzing and designing quantum control systems, as they provide valuable insights into system characteristics, left invertibility and sensitivity.

This paper is organized as follows. Linear quantum systems and system zeros are briefly introduced in Section \ref{Preli}. The special structural properties of linear quantum system in terms of zeros and poles are investigated in Section \ref{ZPLQS}. Section \ref{IZLI} discusses left invertibility of linear quantum systems by means of invariant zeros. A tradeoff in single-input-single-output (SISO) quantum systems is considered in Section \ref{TSQS}. Section \ref{conclu} concludes this paper. 

\emph{Notation.}
$\imath=\sqrt{-1}$ denotes the imaginary unit. For a vector with complex numbers or operators $X=[x_1,\ldots,x_n]^\top$, the complex conjugate of $X$ or its adjoint operator is denoted by $X^\#=[x_1^\ast,\ldots,x_n^\ast]^\top$. Denote  $X^\dagger=(X^\#)^\top$ and  $\breve{X}=\left[\begin{array}{cc}
X^\top & X^\dagger \\   
\end{array}\right]^\top$. Let $J_k={\rm diag}\{I_k,-I_k\}$, define the $\flat$-adjoint of $X\in\mathbb{C}^{2k\times2r}$ by $X^\flat=J_rX^\dagger J_k$. For two matrices $U,V\in\mathbb{C}^{k\times r}$, define the doubled-up matrix as $\Delta(U,V)=\left[\begin{array}{cc}
U & V \\
V^\# & U^\#
\end{array}\right]$.
$\delta_{jk}$ denotes the Kronecker delta function, and $\delta(t-r)$ is the Dirac delta function. Finally, $\otimes$ represents the tensor product.

\section{Preliminaries}\label{Preli}

\subsection{Linear quantum systems}

A linear system model composed of $n$ quantum harmonic oscillators interacting with $m$ input-output boson fields is considered in this paper. The $j$th quantum harmonic oscillator is described by an annihilation operator $a_j$ and its adjoint (creation operator) $a_j^\ast$, and they satisfy the commutation relation $[a_j,a_k^\ast]=\delta_{jk}$, $j,k=1,2,\ldots,n$. Define $a=\left[a_1,\ldots,a_n\right]^\top$. The $j$th input field is represented by an annihilation operator $b_{{\rm in},j}(t)$ and its adjoint $b_{{\rm in},j}^\ast(r)$, which enjoy the following property
\begin{equation*}
[b_{{\rm in},j}(t),b_{{\rm in},k}^\ast(r)]=\delta_{jk}\delta(t-r), ~~ \forall j,k=1,2,\ldots,m, ~~ t,r\in\mathbb{R}.    
\end{equation*}
Denote the integrated annihilation and creation processes by
\begin{equation*}
B_{\rm in}(t)=\int_{-\infty}^t b_{\rm in}(\tau) d\tau, ~~ 
B_{\rm in}^\ast(t)=\int_{-\infty}^t b_{\rm in}^\ast(\tau) d\tau, 
\end{equation*}
where $B_{\rm in}(t)=\left[B_{{\rm in},1}(t),\ldots,B_{{\rm in},m}(t)\right]^\top$.
It is convenient to characterize linear quantum systems by the $(S,\mbf{L},\mbf{H})$ formalism (\cite{GJ09,ZD22}), where $S$ is the scattering operator, the coupling operator between the system and input field is denoted by $\mbf{L}=\left[\begin{array}{cc}
C_- & C_+ \\    
\end{array}\right]\breve{a}$, and  $C_-, C_+\in\mathbb{C}^{m\times n}$. The initial system Hamiltonian $\mbf{H}$ is of the form $\mbf{H}=\frac{1}{2}\breve{a}^\dagger\Omega\breve{a}$, where $\Omega=\Delta(\Omega_-,\Omega_+)$ is Hermitian with $\Omega_-, \Omega+\in\mathbb{C}^{n\times n}$. Assume that $S=I$ (the identity operator). The temporal evolution of a linear quantum system is governed by a unitary operator $U(t,t_0)$, which satisfies the following quantum stochastic differential equation (QSDE)
\begin{equation}\begin{aligned}
dU(t,t_0)=\bigg[&\left(-\imath \mbf{H}-\frac{1}{2}\mbf{L}^\dagger\mbf{L}\right)dt \\
&+dB_{\rm in}^\dagger(t)\mbf{L}-\mbf{L}^\dagger dB_{\rm in}(t)\bigg]U(t,t_0),
\end{aligned}\end{equation}
with the initial condition $U(t,t_0)=I$. In the Heisenberg picture, the evolution of a system operator $\mbf{X}(t)=U^\dagger(t)(\mbf{X}\otimes I)U(t)$ can be expressed as
\begin{equation}\label{eq:X}
\begin{aligned}
d\mbf{X}(t)=&\mathcal{L}_G(\mbf{X}(t))dt+dB_{\rm in}^\dagger(t)[\mbf{X}(t),\mbf{L}(t)] \\
&+[\mbf{L}^\dagger(t),\mbf{X}(t)]dB_{\rm in}(t),
\end{aligned}\end{equation}
where the superoperator 
\begin{equation*}\begin{aligned}
\mathcal{L}_G(\mbf{X}(t))=&-\imath[\mbf{X}(t),\mbf{H}(t)]+\frac{1}{2}\mbf{L}^\dagger(t)[\mbf{X}(t),\mbf{L}(t)] \\
&+\frac{1}{2}[\mbf{L}^\dagger(t),\mbf{X}(t)]\mbf{L}(t).  
\end{aligned}\end{equation*}
The input-output relation is given by
\begin{equation}\label{eq:io}
dB_{\rm out}(t)=\mbf{L}(t)dt+dB_{\rm in}(t).
\end{equation}
More discussions on open quantum systems can be found in, e.g., \cite{GZ00,GJ09,WM10,CKS17,ZD22} and references therein.

Based on Eqs. \eqref{eq:X}-\eqref{eq:io}, the temporal evolution of a linear quantum system in the annihilation-creation operator representation can be described by
\begin{equation}\label{zerosys}\begin{aligned}
\dot{\breve{a}}(t)&=\mathcal{A}\breve{a}(t)+\mathcal{B}\breve{b}_{\rm in}(t), \\
\breve{b}_{\rm out}(t)&=\mathcal{C}\breve{a}(t)+\mathcal{D}\breve{b}_{\rm in}(t), 
\end{aligned}\end{equation}
where the complex-domain system matrices are
\begin{equation}\begin{aligned}
&\mathcal{D}=\Delta(S,0), ~~ \mathcal{C}=\Delta(C_-,C_+), ~~ \mathcal{B}=-\mathcal{C}^\flat \mathcal{D}, \\
&\mathcal{A}=-\imath J_n\Omega-\frac{1}{2}\mathcal{C}^\flat \mathcal{C}.
\end{aligned}\end{equation}
Its corresponding transfer matrix is defined by an {\it irreducible} $G(s)\in \mathbb{C}^{2m\times 2m}$, i.e., each element of $G(s)$ cannot be reducible with a same polynomial of $s$. An alternative real-quadrature representation of the linear quantum system \eqref{zerosys} has been developed in \cite[Theorem 4.4]{ZGPG18} via the Kalman  canonical decomposition, which is
\begin{equation}\label{Kalmansys}\begin{aligned}
\dot{\mbf{x}}(t)&=\bar{A}\mbf{x}(t)+\bar{B}\mbf{u}(t), \\
\mbf{y}(t)&=\bar{C}\mbf{x}(t)+\bar{D}\mbf{u}(t), 
\end{aligned}\end{equation}  where the coordinate transformation \cite[Lemma 4.8]{ZGPG18}  
\begin{equation}\begin{aligned}
&\mbf{x}=\hat{T}^\dagger \breve{a}=\left[\begin{array}{c}
\mbf{q} \\
\mbf{p} 
\end{array}\right], ~ \bar{A}=\hat{T}^\dagger \mathcal{A} \hat{T}, ~ \bar{B}=\hat{T}^\dagger \mathcal{B} V_m^\dagger, ~ \bar{C}=V_m \mathcal{C}\hat{T}, 
\end{aligned}\end{equation}
and
\begin{equation}
\mbf{u}=V_m\breve{b}_{\rm in}=\left[\begin{array}{c}
\mbf{q_{\rm in}} \\
\mbf{p_{\rm in}} 
\end{array}\right], ~ \mbf{y}=V_m\breve{b}_{\rm out}=\left[\begin{array}{c}
\mbf{q_{\rm out}} \\
\mbf{p_{\rm out}} 
\end{array}\right], ~ \bar{D}=V_m \mathcal{D} V_m^\dagger,
\end{equation}
with the unitary matrix $V_m=\frac{1}{\sqrt{2}}\left[\begin{array}{cc}
I_m & I_m \\
-\imath I_m & \imath I_m
\end{array}\right]$. In the sequel, we denote the transfer matrix of system \eqref{Kalmansys} by $\mathbb{G}(s)$. 



\subsection{Invariant zeros and transmission zeros}

According to \cite[Definition 3.16]{ZDG96}, the {\it invariant zeros of the linear quantum system realization}  \eqref{Kalmansys} are the complex numbers $s_0$, which satisfy the following inequality
\begin{equation}\label{eq:inv_zeros}
{\rm{rank}}\ P(s_0)<{\rm normalrank}\ P(s),
\end{equation}
where 
\begin{equation}  \label{eq:P(s)}
P(s)\triangleq \left[\begin{array}{cc}\bar{A}-sI & \bar{B} \\
\bar{C} & \bar{D}\end{array}\right],
\end{equation}
and the normalrank is defined by \cite[Definition 3.11]{ZDG96}.
\begin{remark}\label{2n2m}
It is worth mentioning that ${\rm normalrank}\ P(s)=2(n+m)$ always holds for open linear quantum systems \eqref{Kalmansys}  due to the  orthogonality of $\bar{D}$.
\end{remark} 

\bmrk The following is given  in \cite{ZDG96}.
(Notice that $P(s)$ is a square matrix of $2(n+m)$.) Let $s_0$ be an invariant zero.  Then there is  $0\neq x\in\mathbb{C}^{2n}$ and $u\in\mathbb{C}^{2m}$ such that
\beq \la{eq:inv_zero}
P(s_0)\left[
\bey{c}
x\\
u
\eey
\right] = \left[
\bey{c}
(\bar{A}-s_0I)x+\bar{B}u\\
\bar{C}x+\bar{D}u
\eey
\right] =0.
\eeq
If $u=0$, then $s_0$ is an unobservable mode.  On the other hand, there is $0\neq u\in\mathbb{C}^{2n}$ and $v\in\mathbb{C}^{2m}$ such that
\beq \la{eq:inv_zero_2}
[y^\dag ~~ v^\dag]P(s_0) = [y^\dag (\bar{A}-s_0I) + v^\dag \bar{C} ~~  y^\dag \bar{B} +  v^\dag\bar{D} ]  =0.
\eeq
If $v=0$, then $s_0$ is an uncontrollable mode.
\emrk

As in the classical realm, by decomposition into observable
and unobservable eigenvalues,  $P(s)$  can be rewritten as 
\begin{equation}\label{Feb3-2}
\tilde{P}(s)=\left[\begin{array}{ccc}
\bar{A}_o-sI_{2n-r} & 0 & \bar{B}_o  \\
\bar{A}_c & \bar{A}_{\bar{o}}-sI_{r} & \bar{B}_{\bar{o}} \\
\bar{C}_o & 0 & \bar{D}
\end{array}\right],    
\end{equation}
where 
the observable and unobservable eigenvalues are the eigenvalues of $\bar{A}_o\in\mathbb{R}^{(2n-r)\times(2n-r)}$ and $\bar{A}_{\bar{o}}\in\mathbb{R}^{r\times r}$, respectively.  Here, $r$ is the dimension of the unobservable subspace.  

Thus, the invariant zeros of the linear quantum system realization \eqref{Kalmansys} consist of the eigenvalues of $\bar{A}_{\bar{o}}$ (which are commonly called the unobservable eigenvalues), and the invariant zeros of 
the observable subsystem realization or equivalently
\begin{equation}
P_o(s)=\left[\begin{array}{cc}
\bar{A}_o-sI_{2n-r} & \bar{B}_o \\
\bar{C}_o & \bar{D}
\end{array}\right].    
\end{equation}

The following result is recalled to prepare the definition of transmission zeros of linear systems.

\begin{lemma}\cite[Theorem 2.3]{Maciejowski89}\label{McMillan}
Let $G(s)$ be a rational matrix of normal rank $r$. Then $G(s)$ may be transformed by a series of elementary row and column operations into a pseudo-diagonal (can be nonsquare) rational matrix $M(s)$ of the  form
\begin{equation}\label{Eq. Smith-McMillan}\begin{aligned}
M(s)={\rm diag}\left\{\frac{\alpha_1(s)}{\beta_1(s)},\frac{\alpha_2(s)}{\beta_2(s)},\cdots,\frac{\alpha_r(s)}{\beta_r(s)},0,\cdots,0\right\},
\end{aligned}\end{equation}
in which the monic polynomials $\{\alpha_i(s),\beta_i(s)\}$ are coprime for each $i$ and satisfy the divisibility properties
\begin{equation}
\alpha_i(s)|\alpha_{i+1}(s), ~~ \beta_{i+1}(s)|\beta_i(s), ~~ i=1,\ldots,r-1,   
\end{equation}
In the literature, $M(s)$ is commonly referred to as the Smith-McMillan form of $G(s)$.  
\end{lemma}

In the above lemma, the Smith-McMillan form $M(s)$ is derived by a series of elementary row and column operations on the transfer matrix $G(s)$, which can be also equivalently written as $M(s)=U(s)G(s)V(s)$, where $U(s)$, $V(s)$ are polynomial unimodular matrices. 


There are various definitions of transmission zeros in the literature; see for example \cite{kailath1980,DCC1982,CD1991,ZDG96}. In this paper, we adopt the definition of transmission zeros used in \cite[Definition 3.14]{ZDG96}. That is, the {\it transmission zeros} of the transfer function matrix $G(s)$ are the roots of all the numerator polynomials of the Smith-McMillan form given in Lemma \ref{McMillan}.  A complex number $s_0\in\mathbb{C}$ is called a {\it pole} of the transfer function matrix $G(s)$ if it is a root of any one of the polynomials $\beta_j(s)$ in the Smith-McMillan \eqref{Eq. Smith-McMillan} form of $G(s)$. 


\begin{remark}
By definition, transmission zeros are defined in terms of transfer functions, while invariant zeros are defined in terms of state-space realizations. However, by  \cite[Corollary 3.35]{ZDG96} a transmission zero must be an invariant zero, and moreover by \cite[Theorem 3.34]{ZDG96} the transmission zeros and invariant zeros are identical for system minimal realizations.
\end{remark}

The following lemma presents a useful  criterion for the transmission zero if it is not a pole.  
 
\begin{lemma} \label{lem:C330}\cite[Corollary 3.30]{ZDG96}
Let $G(s)$ be a square $2m\times 2m$ proper transfer matrix and ${\rm det}[G(s)]\not\equiv0$. Suppose $s_0\in\mathbb{C}$ is not a pole of $G(s)$. Then $s_0\in\mathbb{C}$ is a transmission zero of $G(s)$ if and only if ${\rm det}[G(s_0)]=0$.
\end{lemma}

\bmrk
Clearly, for open linear quantum systems,  ${\rm det}[G(s)]\not\equiv0$ always holds as the $D$-matrix is unitary.
\emrk

\section{Zeros and poles of linear quantum  systems}\label{ZPLQS}

In this section, we study the relation between zeros and poles for linear quantum systems.

We begin with the following result.

\begin{proposition}\label{prop_G and inv G}
$s_0$ is a pole if and only if $-s_0^\ast$ is a transmission zero  of $G(s)$.
\end{proposition}

\emph{Proof.} By the Smith-McMillan form \eqref{Eq. Smith-McMillan}, $s_0$ being a pole of $G(s)$ implies that there exists at least a polynomial $\beta_i(s)$ satisfying $\beta_i(s_0)=0$, and thus $s_0$ is a transmission zero of $G^{-1}(s)$.  As $G^{-1}(s)=G(-s^\ast)^\flat$ for linear quantum systems \cite{GJN10,ZJ13}, $s_0$ is also a transmission zero of $G(-s^\ast)^\flat$.  
By \cite[Lemma 3.26]{ZDG96},
\begin{equation}\label{Feb24-1}
U(s)G(s)V(s)=M(s)=
{\rm diag}\left\{\frac{\alpha_1(s)}{\beta_1(s)},\frac{\alpha_2(s)}{\beta_2(s)},\cdots,\frac{\alpha_{2m}(s)}{\beta_{2m}(s)}\right\},
\end{equation}
yields that 
\begin{equation}\begin{aligned}
&V(-s^\ast)^\flat G(-s^\ast)^\flat U(-s^\ast)^\flat=M(-s^\ast)^\flat \\
=\ &{\rm diag}\left\{\frac{\alpha_1^\ast(-s^\ast)}{\beta_1^\ast(-s^\ast)},\frac{\alpha_2^\ast(-s^\ast)}{\beta_2^\ast(-s^\ast)},\cdots,\frac{\alpha_{2m}^\ast(-s^\ast)}{\beta_{2m}^\ast(-s^\ast)}\right\},  
\end{aligned}\end{equation}
in which there must be a polynomial $\alpha_j^\ast(-s_0^\ast)=0$ and thus $\alpha_j(-s_0^\ast)=0$ in \eqref{Feb24-1}. As a result, $-s_0^\ast$ is a transmission zero. The converse of Proposition \ref{prop_G and inv G} can be proved in a similar way.  $\Box$

\begin{example}
Consider a classical linear system with system matrices $A=B=C=I_2$ and $D=\mbf{0}$. 
It can be calculated that the transfer function is $G(s)=\frac{1}{s-1}I_2
$. Clearly, the pole of $G(s)$ is $1$, while there is no transmission zero. Thus, Proposition \ref{prop_G and inv G} in general does not hold for classical linear systems. 
\end{example}

\begin{remark} \label{rem:imaginary}
According to Proposition \ref{prop_G and inv G}, a purely imaginary pole is also a purely imaginary transmission zero, and vice versa.
\end{remark}




Further demonstration of the relation between invariant zero and pole for linear quantum systems is given by the following proposition.

\bprop\label{Feb3-3} 
If $s_0$ is an eigenvalue of $\mathcal{A}$, then $-s_0^\ast$ must be an invariant zero of the linear quantum system realization \eqref{zerosys}, and vice versa.
\eprop

\emph{Proof.} $s_0$ is an eigenvalue of $\mathcal{A}$ if and only if $s_0^\ast$ is an eigenvalue of $\mathcal{A}^\dagger$. Notice that
\begin{equation}\label{RSM2}
{\rm det}\left[s_0^\ast I-\mathcal{A}^\flat\right]=
{\rm det}\left[s_0^\ast I-J_n \mathcal{A}^\dagger J_n\right]
= {\rm det} \left[s_0^\ast I-\mathcal{A}^\dagger\right]=0.
\end{equation}
 Thus $s_0^\ast$ is also an eigenvalue of $\mathcal{A}^\flat$. Since $\mathcal{D}$ is unitary, it is easy to verify that the following identity holds
\begin{equation}\label{RSM1}\begin{aligned}
\left[\begin{array}{cc}\mathcal{A}-sI & \mathcal{B} \\ \mathcal{C} & \mathcal{D}\end{array}\right]
\left[\begin{array}{cc}I & 0 \\ -\mathcal{D}^{-1}\mathcal{C} & I\end{array}\right]=
\left[\begin{array}{cc}\mathcal{A}-sI+\mathcal{C}^\flat\mathcal{C} & \mathcal{B} \\ 0 & \mathcal{D}\end{array}\right],
\end{aligned}\end{equation}
where the physical realizability condition of linear quantum systems $\mathcal{B}=-\mathcal{C}^\flat\mathcal{D}$ has been used in the derivation. By \eqref{RSM1}, 
\begin{equation}\label{Feb3-1}\begin{aligned}
&{\rm det}[P(s)]={\rm det}\left[\mathcal{A}-sI+\mathcal{C}^\flat\mathcal{C}\right] \\
=\ &{\rm det}\left[-sI-\imath J_n\Omega+\frac{1}{2}\mathcal{C}^\flat\mathcal{C}\right]={\rm det}\left[sI+\mathcal{A}^\flat\right].
\end{aligned}\end{equation}
Let $s=-s_0^\ast$ in \eqref{Feb3-1}, by \eqref{RSM2} we have ${\rm det}[P(-s_0^\ast)]=0$, which means that $-s_0^\ast$ must be an invariant zero of the linear quantum system realization \eqref{zerosys}. Conversely, if $s_0$ is an invariant zero of the linear quantum system realization \eqref{zerosys}, then by \eqref{Feb3-1} ${\rm det}[P(s_0)]=0$ implies $s_0$ is an eigenvalue of $-\mathcal{A}^\flat$. Thus, $-s_0^\ast$ is an eigenvalue of $\mathcal{A}$.  $\Box$


\begin{example}
Consider a classical linear system with system matrices $A=\left[\begin{array}{cc}
1 & 0 \\
0 & 2
\end{array}\right]$, $B=C=\left[\begin{array}{cc}
1 & 0 \\
0 & 0
\end{array}\right]$, and $D=I_2$. It can be calculated that the Smith-McMillan form of the  transfer matrix $G(s)=\left[\begin{array}{cc}
\frac{s}{s-1} & 0 \\
0 & 1
\end{array}\right]$ is 
$M(s)=\left[\begin{array}{cc}
\frac{1}{s-1} & 0 \\
0 & s
\end{array}\right]$. Thus, the pole and transmission zero of this system are $1$ and $0$, respectively. However, the eigenvalues of $A$ are $1$ (also the pole of $G(s)$) and $2$, while the invariant zeros of this system realization $P(s)$ are $0$ and $2$, which indicates that Proposition \ref{Feb3-3} in general does not  hold only for classical linear systems.
\end{example}

\begin{remark}
Obviously, the set of poles is a subset of eigenvalues of $\mathcal{A}$, while the set of transmission zeros is a subset of invariant zeros of system realizations. Loosely speaking, Proposition \ref{Feb3-3} is a generalization of Proposition \ref{prop_G and inv G}. By Propositions \ref{prop_G and inv G}-\ref{Feb3-3}, there exist a one-to-one correspondence between  poles and transmission zeros of a transfer function and another one-to-one correspondence between eigenvalues of $\mathcal{A}$ and invariant zeros of a system realizations.   
\end{remark}

Before going into the main result, Theorem \ref{Feb3-4},  of this section, we first present an assumption, which is a necessary condition for Theorem \ref{Feb3-4}.

\begin{assumption}\label{Feb-h}
Based on the Kalman canonical form derived in \cite[Eq. (4)]{ZPL20}, assume further the poles of ``$h$" (controllable and unobservable plus uncontrollable and observable) and $A_{\bar{c}\bar{o}}$ (uncontrollable and unobservable) subsystems are purely imaginary.
\end{assumption}


\begin{remark}
Clearly, Assumption \ref{Feb-h} holds for linear  passive quantum systems. Actually, many physical systems  satisfy Assumption \ref{Feb-h}; see for example \cite{DFK+12,NY14,LOW+21}. 
\end{remark}

The following theorem illustrates the relation between invariant zeros of linear quantum systems and the eigenvalues of $\mathcal{A}$.

\begin{theorem}\label{Feb3-4}
The set of invariant zeros of the linear quantum system realization \eqref{zerosys} is a union of the set of $\mathcal{A}_{\bar{o}}$  eigenvalues and the set of negative conjugate of $\mathcal{A}_o$  eigenvalues under Assumption \ref{Feb-h}.
\end{theorem}

\emph{Proof.} By Proposition \ref{Feb3-3}, the invariant zeros of the linear quantum system realization \eqref{zerosys} are the eigenvalues of $-\mathcal{A}^\flat$. As the set of eigenvalues of $\mathcal{A}$ can be expressed as a union of the set of observable eigenvalues $\lambda(\mathcal{A}_o)$ and the set of unobservable eigenvalues $\lambda(\mathcal{A}_{\bar{o}})$, it can be verified that the set of eigenvalues of $-\mathcal{A}^\flat$ can be also expressed as the union of $-\lambda^\ast(\mathcal{A}_o)$ and $-\lambda^\ast(\mathcal{A}_{\bar{o}})$ (or equivalently, $\lambda(\mathcal{A}_{\bar{o}})$) by Assumption \ref{Feb-h}. Thus, by Proposition \ref{Feb3-3}, the set of invariant zeros of the linear quantum system realization \eqref{zerosys} is $-\lambda^\ast(\mathcal{A}_o)\cup\lambda(\mathcal{A}_{\bar{o}})$. $\Box$

\section{Invariant zeros and left invertibility}\label{IZLI}

As an isolated quantum system evolves unitarily, its temporal dynamics is invertible. For an open quantum linear system $G$, it is shown in \cite{GJN10,ZJ13} that its inverse $G^{-1}$ always exists. Unfortunately,   $G^{-1}$ is unstable if $G$ is stable. In the classical (non-quantum mechanical) control literature, left invertibility is a critical concept in the study of linear systems \cite{Moylan77}. For example, it is proved in \cite{DD2023} that if a linear system is left invertible, then there exists a  stable inversion such that the input to the original system  can be reconstructed from the output. In this section, we study left invertibility and extend it to the linear quantum realm. 

We first recall the definitions of left invertibility for linear systems.


\begin{definition}\cite[Definition 5]{DD2023}
A finite-dimensional linear time-invariant (FDLTI) system 
\begin{equation*}\begin{aligned}
\dot{x}(t)&=Ax(t)+Bu(t), \\
y(t)&=Cx(t)+Du(t)
\end{aligned}\end{equation*}
is said to be:

\begin{itemize}

    \item strongly-(s.-)left invertible if for any initial condition $x(0)$
\begin{equation}
y(t)=0 ~~ \text{for} ~~ t>0 ~~~~ \Longrightarrow ~~~~ u(t)=0 ~~ \text{for} ~~ t>0; 
\end{equation}
    
    \item asymptotically strongly-(a.s.-)left invertible if for any initial condition $x(0)$
\begin{equation}
y(t)=0 ~~ \text{for} ~~ t>0 ~~~~ \Longrightarrow ~~~~ u(t)\longrightarrow0 ~~ \text{as} ~~ t\longrightarrow \infty;  
\end{equation}
    
    \item asymptotically $\text{strong}^{\star}$ ($\text{a.s.}^{\star}$-)left invertible if for any initial condition $x(0)$
\begin{equation}
y(t)\longrightarrow0 ~~ \text{as} ~~ t\longrightarrow\infty ~~~~ \Longrightarrow ~~~~ u(t)\longrightarrow0 ~~ \text{as} ~~ t\longrightarrow \infty.  
\end{equation}
    
\end{itemize}
\end{definition}

Due to the special structure of linear quantum systems, we have the following results concerning the relation between their invariant zeros and left invertibility.

\begin{theorem}\label{Thm6.1}
The linear quantum system \eqref{zerosys} under  Assumption \ref{Feb-h} is a.s.-left invertible if and only if the set of $\mathcal{A}_o$ eigenvalues is in the right half of the complex plane. 
\end{theorem}

\emph{Proof.} By \cite[Theorem 14]{DD2023} and Remark \ref{2n2m}, a linear quantum   system  is a.s.-left invertible if and only if the subset of invariant zeros of $P(s)$ that do not belong to the set of $(\mathcal{A},\mathcal{C})$-unobservable eigenvalues is in the left half of the complex plane. By Theorem \ref{Feb3-4}, the latter is equivalent to the negative conjugates of $\mathcal{A}_o$ eigenvalues are in the left half of the complex plane, which means all the eigenvalues of $\mathcal{A}_o$ are in the right half of the complex plane. $\Box$

The following example demonstrates that Assumption \ref{Feb-h} is indispensable for the validity of Theorem \ref{Thm6.1}.
\begin{example}\la{eq:Example_h}
Consider a linear quantum system with $A=\left[\begin{array}{cc}
-1 & 0 \\
0 & 1
\end{array}\right]$, $B=C=\left[\begin{array}{cc}
0 & 1 \\
0 & 0
\end{array}\right]$, and $D=I_2$.  This quantum system is physically realizable since $A+A^\sharp+BB^\sharp=\mbf{0}$, $B=-C^\sharp D$ (\cite{JNP08,ZGPG18,ZD22}), and is actually an ``$h$" system with system Hamiltonian $\mbf{H}=\frac{1}{2}x^\top H x$ and coupling operator $\mbf{L}=\Gamma x$, where $H=\left[\begin{array}{cc}
0 & -1 \\
-1 & 0
\end{array}\right]$ and $\Gamma=\left[\begin{array}{cc}
0 & \frac{1}{\sqrt{2}} 
\end{array}\right]$. 
It is straightforward that $y_2(t)=u_2(t)$ and $y_1(t)=x_2(t)+u_1(t)$. If $y_2=0$ for $t>0$, then $u_2=0$ for $t>0$, which indicates that the second output is a.s.-left invertible (even s.-left invertible). However, if $y_1=0$ for $t>0$, then $u_1(t)=-x_2(t)=-e^tx_2(0)$, which is divergent for any nonzero initial state $x_2(0)$. Thus, this quantum system is not a.s.-left invertible.
\end{example}

\begin{remark}
Compared with \cite[Theorem 14]{DD2023}, Theorem \ref{Thm6.1} is derived with the aid of the special linear quantum systems structure. That is, the equivalent condition is not applied to linear classical systems. Consider a classical system with system matrices $A=\left[\begin{array}{cc}
1 & 0 \\
0 & 0
\end{array}\right]$, $B=\left[\begin{array}{c}
-1 \\
0
\end{array}\right]$, $C=\left[\begin{array}{cc}
1 & 0  
\end{array}\right]$, $D=1$. Clearly, this system has a $co$ subsystem and a $\bar{c}\bar{o}$ subsystem and Assumption \ref{Feb-h} holds.  The observable eigenvalue is $1$, which  satisfies the condition ``the set of $\mathcal{A}_o$ eigenvalues is in the right half of the complex plane" given in Theorem \ref{Thm6.1}. However, $y=0$ implies that $u(t)=-e^{2t}u_0$ ($u_0$ is a constant), which does not tend to zero. Thus, this system is not a.s.-left invertible. 
\end{remark}

\bmrk
As mentioned at the beginning of this section, $G^{-1}$ is unstable if the linear quantum system $G$ is stable; or equivalently, $G$ is unstable if $G^{-1}$ is stable. However, an unstable system may have stable modes.  
Theorem \ref{Thm6.1} provides a condition for the left invertibility of a linear quantum system by exploring its inner structure. 
\emrk

We conclude this section with a final remark.

\begin{remark}
By \cite[Theorem 15]{DD2023} and Remark \ref{2n2m}, it is straightforward to obtain that 
the a.s.-left invertibility and $\text{a.s.}^{\star}$-left invertibility are equivalent to each other for linear quantum systems.
\end{remark}

\section{A tradeoff in SISO quantum  linear systems}\label{TSQS}

\begin{figure}[thpb]
      \centering
      \includegraphics[scale=0.58]{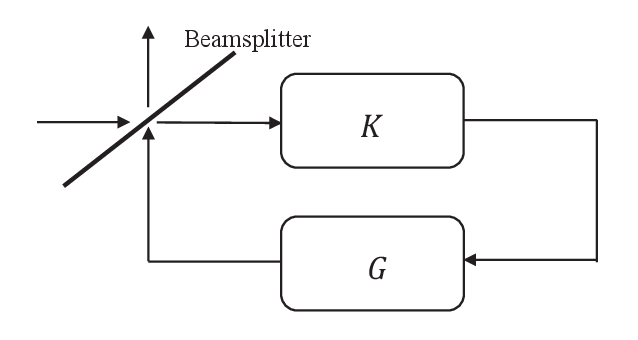}
      \caption{A quantum coherent feedback network composed of two linear quantum systems $G$, $K$, and a beamsplitter.}
      \label{fig_LFT}
   \end{figure}

In this section, we study the tradeoff of a coherent feedback network in Fig. \ref{fig_LFT} where both $G$ and $K$ are single-input-single-output  (SISO) linear quantum systems.



Assume that  $\Omega$ is purely imaginary and $\mathcal{C}$ is real or purely imaginary for the SISO linear quantum system $G$. Then its transfer matrix associated with the Kalman canonical form \eqref{Kalmansys} is of a diagonal form
\begin{equation}\label{Mar20-1}
\mathbb{G}(s)={\rm diag}\{\mathbb{G}_q(s),\mathbb{G}_p(s)\}.  
\end{equation}
If $\Omega_-=0$,
that is, the system  and the input field are resonant, then it is easy to show that the transfer matrix \eqref{Mar20-1} satisfies
\begin{equation}\label{tradeoffG}
\mathbb{G}_q(s)\mathbb{G}_p(-s)=1.
\end{equation}

Under this circumstance, the transfer matrix reduces to 
\begin{equation}\begin{aligned}
\mathbb{G}(s)=
\left[\begin{array}{cc}
\frac{s+\imath\Omega_+-\frac{1}{2}\mathbb{C}_q\mathbb{C}_p}{s+\imath\Omega_++\frac{1}{2}\mathbb{C}_q\mathbb{C}_p} & 0 \\
0 & \frac{s-\imath\Omega_+-\frac{1}{2}\mathbb{C}_q\mathbb{C}_p}{s-\imath\Omega_++\frac{1}{2}\mathbb{C}_q\mathbb{C}_p}
\end{array}\right],
\end{aligned}\end{equation}
where $\mathbb{C}_q=C_-+C_+$, $\mathbb{C}_p=C_--C_+$. Moreover,  if $\Omega_+=\pm\frac{\imath}{2}\mathbb{C}_q\mathbb{C}_p$, then either   $\mathbb{G}_q(s)$ has a zero and $\mathbb{G}_p(s)$ has a pole or $\mathbb{G}_q(s)$ has a pole and $\mathbb{G}_p(s)$ has a zero at the origin. This means that one quadrature output reduces the input noise and asymptotically produces ideal squeezing at low frequencies. However, the other quadrature output is unstable and diverges to infinity as the frequency tends to $0$. 

\begin{example}\cite{ZJ11}
Consider a degenerate parametric amplifier (DPA) with $\Omega_-=0$, $\Omega_+=\frac{\imath\epsilon}{2}$, $C_-=\sqrt{\kappa}$, and $C_+=0$, $\epsilon<\kappa$. It can be directly calculated that the transfer matrix is given by
\begin{equation}
\mathbb{G}(s)=
\left[\begin{array}{cc}
\frac{s-\frac{\epsilon}{2}-\frac{\kappa}{2}}{s-\frac{\epsilon}{2}+\frac{\kappa}{2}} & 0 \\
0 & \frac{s+\frac{\epsilon}{2}-\frac{\kappa}{2}}{s+\frac{\epsilon}{2}+\frac{\kappa}{2}}
\end{array}\right],
\end{equation}
In the limit $\epsilon=\pm\kappa$, there is a zero and pole at the origin.
\end{example}

For the coherent feedback network in Fig. \ref{fig_LFT}, let $\Omega$, $\mathcal{C}$, and $\Omega^\prime$, $\mathcal{C}^\prime$ be the system parameters of plant $G$ and controller $K$, respectively.  Assume  the plant $G$ satisfies \eqref{tradeoffG}, and the controller $K$ satisfies $\mathbb{K}_q(s)\mathbb{K}_p(-s)=1$. That is, $\mathcal{C}$ and $\mathcal{C}^\prime$ are real or purely imaginary, both $\Omega$ and $\Omega^\prime$ are purely imaginary with $\Omega_-=\Omega_-^\prime=0$. The beamsplitter is of the form $\left[\begin{array}{cc}
\alpha & \beta \\
\beta & -\alpha
\end{array}\right]$, with real parameters $\alpha$, $\beta$, s.t. $\alpha^2+\beta^2=1$. The transfer matrix for the coherent feedback network is given by $\mathbb{T}(s)=
\left[\begin{array}{cc}
\mathbb{T}_q(s) & 0 \\
0 & \mathbb{T}_p(s)
\end{array}\right]$, where
\begin{equation}\label{Mar15-6}\begin{aligned}
\mathbb{T}_q(s)=\frac{\alpha+\mathbb{G}_q\mathbb{K}_q}{1+\alpha\mathbb{G}_q\mathbb{K}_q}, ~~ 
\mathbb{T}_p(s)=\frac{\alpha+\mathbb{G}_p\mathbb{K}_p}{1+\alpha\mathbb{G}_p\mathbb{K}_p}.
\end{aligned}\end{equation}
 It can be verified that $\mathbb{T}_q(s)\mathbb{T}_p(-s)=1$, which means the structure \eqref{tradeoffG} is preserved; see also \cite{Yan07}. 

Suppose that $\alpha+\mathbb{G}_q\mathbb{K}_q=0$, then we have
\begin{equation}\label{Mar15-2}\begin{aligned}
\alpha+\frac{s+\imath\Omega_+-\frac{1}{2}\mathbb{C}_q\mathbb{C}_p}{s+\imath\Omega_++\frac{1}{2}\mathbb{C}_q\mathbb{C}_p}    \frac{s+\imath\Omega_+^\prime-\frac{1}{2}\mathbb{C}_q^\prime\mathbb{C}_p^\prime}{s+\imath\Omega_+^\prime+\frac{1}{2}\mathbb{C}_q^\prime\mathbb{C}_p^\prime}=0.
\end{aligned}\end{equation}
If $\mathbb{T}_q(s)$ has a zero at the origin, then by \eqref{Mar15-2} we have
\begin{equation}\label{Mar15-3}\begin{aligned}
&(1+\alpha)\left(\frac{1}{4}\mathbb{C}_q\mathbb{C}_p\mathbb{C}_q^\prime\mathbb{C}_p^\prime-\Omega_+\Omega_+^\prime\right) \\
-\ &(1-\alpha)\left(\frac{\imath}{2}\mathbb{C}_q\mathbb{C}_p\Omega_+^\prime+\frac{\imath}{2}\mathbb{C}_q^\prime\mathbb{C}_p^\prime\Omega_+\right)=0,
\end{aligned}\end{equation}
which demonstrates that  $\mathbb{T}_q(s)$ can produce ideal squeezing at low frequencies. Similarly, it can be verified that $\mathbb{T}_p(s)$ producing ideal squeezing at low frequencies requires that
\begin{equation}\label{Mar15-4}\begin{aligned}
&(1+\alpha)\left(\frac{1}{4}\mathbb{C}_q\mathbb{C}_p\mathbb{C}_q^\prime\mathbb{C}_p^\prime-\Omega_+\Omega_+^\prime\right) \\
+\ &(1-\alpha)\left(\frac{\imath}{2}\mathbb{C}_q\mathbb{C}_p\Omega_+^\prime+\frac{\imath}{2}\mathbb{C}_q^\prime\mathbb{C}_p^\prime\Omega_+\right)=0.
\end{aligned}\end{equation}
As a result, the coherent feedback network can realize ideal squeezing under the condition \eqref{Mar15-3} or \eqref{Mar15-4}, even if the original plant $G$ cannot, i.e., $\Omega_+\neq\pm\frac{\imath}{2}\mathbb{C}_q\mathbb{C}_p$.

The following two examples indicate that the coherent feedback network can realize the ideal input squeezing by means of a passive controller or a active controller with the same system parameter $\mathcal{C}$ as the plant.

\begin{example} We first look at the setup in \cite{GW09}, where $K$ is 1. Then the constraints in \eqref{Mar15-3} and \eqref{Mar15-4} reduce to
\begin{equation}\label{Mar20-11}
\frac{1+\alpha}{2}\mathbb{C}_q\mathbb{C}_p\mp\imath(1-\alpha)\Omega_+=0.
\end{equation}
If $\alpha=\frac{\pm\imath\Omega_+-\frac{1}{2}\mathbb{C}_q\mathbb{C}_p}{\pm\imath\Omega_++\frac{1}{2}\mathbb{C}_q\mathbb{C}_p}$,  then \eqref{Mar20-11} holds. Thus the coherent feedback network in Fig. \ref{fig_LFT} realizes ideal input squeezing.
\end{example}

\begin{example}
If the desired controller $K$ has the same system parameter $\mathcal{C}$ as the plant $G$, i.e., $\mathbb{C}_q^\prime=\mathbb{C}_q$ and $\mathbb{C}_p^\prime=\mathbb{C}_p$, then the constraints in \eqref{Mar15-3} and \eqref{Mar15-4} reduce to
\begin{equation}\label{Mar15-9}\begin{aligned}
(1+\alpha)\left(\frac{1}{4}\mathbb{C}_q^2\mathbb{C}_p^2-\Omega_+\Omega_+^\prime\right)\mp\frac{\imath(1-\alpha)}{2}\mathbb{C}_q\mathbb{C}_p(\Omega_++\Omega_+^\prime)=0,
\end{aligned}\end{equation}   
which yields the system parameter of $K$ 
\begin{equation}\label{Mar15-10}
\Omega_+^\prime=\frac{\mp\imath\mathbb{C}_q\mathbb{C}_p}{2}
\frac{(1+\alpha)\mathbb{C}_q\mathbb{C}_p\mp2(1-\alpha)\imath\Omega_+}{(1-\alpha)\mathbb{C}_q\mathbb{C}_p\mp2(1+\alpha)\imath\Omega_+}.
\end{equation}
Thus, the coherent feedback network including controller $K$ with the same system parameter $\mathcal{C}$ as the plant $G$ and the pump $\Omega_+^\prime$ given by \eqref{Mar15-10} can realize ideal input squeezing. 
\end{example}

By definition \cite{SBG12}, the sensitivity function of the coherent feedback network  can be calculated in the following way
\begin{equation}
\frac{d\mathbb{T}_j}{d\mathbb{G}_j}=\mathtt{S}_j\frac{\mathbb{T}_j}{\mathbb{G}_j}, ~~ j=q,p,    
\end{equation}
where the sensitivity function 
\begin{equation}
\mathtt{S}=\left[\begin{array}{cc}
\mathtt{S}_q & 0 \\
0 & \mathtt{S}_p
\end{array}\right]
=\frac{\beta^2\mathbb{G}\mathbb{K}}{(I+\alpha\mathbb{G}\mathbb{K})(\alpha I+\mathbb{G}\mathbb{K})}.    
\end{equation}
If the coherent feedback network  is designed to realize input squeezing at low frequencies, then $\alpha+\mathbb{G}_j\mathbb{K}_j\rightarrow0$ in \eqref{Mar15-6} implies the corresponding sensitivity function $\mathtt{S}_j$ diverges to infinity, $j=q,p$. Thus, the coherent feedback network  is extremely sensitive to the uncertainty of the plant $G$; see also \cite[Fig. 1]{Yan07}. This is a fundamental tradeoff between squeezing and system robustness posed by the zeros of the coherent feedback network.






\addtolength{\textheight}{-3cm}   




\section{Conclusion}\label{conclu}

In this paper, transmission zeros, invariant zeros and poles have been  investigated in the ream of linear quantum systems. They provide valuable insights into system characteristics and physical properties, e.g., left invertibility, input squeezing, which are crucial for various applications in quantum computing and quantum communication. A thorough understanding of these concepts  in the realm of quantum control systems is our future research endeavor.









\bibliographystyle{IEEEtran}
\bibliography{gzhang.bib}

\end{document}